\documentclass[a4paper,fleqn]{cas-sc}
\usepackage[version=4]{mhchem}
\usepackage{amsmath}
\usepackage{hyperref}
\usepackage[authoryear,longnamesfirst]{natbib}
\usepackage{adjustbox}
\def\tsc#1{\csdef{#1}{\textsc{\lowercase{#1}}\xspace}}
\tsc{WGM}
\tsc{QE}
\tsc{EP}
\tsc{PMS}
\tsc{BEC}
\tsc{DE}

\begin{document}
	\let\WriteBookmarks\relax
	\def\floatpagepagefraction{1}
	\def\textpagefraction{.001}
	\shorttitle{Methylamine towards G358.93--0.03 MM1}
	\shortauthors{Manna \& Pal}
	\title [mode = title]{Detection of possible glycine precursor molecule methylamine towards the hot molecular core G358.93--0.03 MM1}  
	
	\author[1]{Arijit Manna}
	\ead{arijitmanna@mcconline.org.in, tel: +918777749445}
	\address[1]{Department of Physics and Astronomy, Midnapore City College, Kuturia, Bhadutala, Paschim Medinipur, West Bengal, India 721129}
	\cormark[1]
	\author[1]{Sabyasachi Pal}
	
	\begin{abstract}
The search for the simplest amino acid, glycine (\ce{NH2CH2COOH}), in the interstellar medium (ISM) has become a never-ending story for astrochemistry and astrophysics researchers because that molecule plays a possible connection between the Universe and the origin of life. In the last forty years, all searches for \ce{NH2CH2COOH} in the ISM at millimeter and submillimeter wavelengths have failed. Since the detection of \ce{NH2CH2COOH} in the ISM was extremely difficult, we aimed to search for the possible precursors of \ce{NH2CH2COOH}. Earlier, many laboratory experiments have suggested that methylamine (\ce{CH3NH2}) plays an important role in the ISM as a possible precursor of \ce{NH2CH2COOH}. After spectral analysis using the local thermodynamic equilibrium (LTE) model, we identified the rotational emission lines of \ce{CH3NH2} towards the hot molecular core G358.93--0.03 MM1 using the Atacama Large Millimeter/Submillimeter Array (ALMA). The column density of \ce{CH3NH2} towards the G358.93--0.03 MM1 was estimated to be (1.10$\pm$0.31)$\times$10$^{17}$ cm$^{-2}$ with an excitation temperature of 180.8$\pm$25.5 K. The fractional abundance of \ce{CH3NH2} with respect to \ce{H2} towards the G358.93--0.03 MM1 was (8.80$\pm$2.60)$\times$10$^{-8}$. The column density ratio of \ce{CH3NH2} and \ce{NH2CN} towards G358.93--0.03 MM1 was (1.86$\pm$0.95)$\times$10$^{2}$. The estimated fractional abundance of \ce{CH3NH2} towards the G358.93--0.03 MM1 agrees fairly well with the previous three-phase warm-up chemical modelling abundance of \ce{CH3NH2}. We also discussed the possible formation mechanism of \ce{CH3NH2}, and we find that \ce{CH3NH2} is most probably formed via the reactions of radical \ce{CH3} and radical \ce{NH2} on the grain surface of G358.93--0.03 MM1.
	\end{abstract}
	 
	\maketitle
	
\section{Introduction}
\label{sec:intro} 
The initial chemical development leading to the formation of life is believed to have formed in molecular clouds, proceeded inside the protoplanetary disk, and transported into the planetary atmosphere with the help of comets and asteroids \citep{suz19}. The investigation of complex nitrogen-bearing molecules like hydrogen cyanide (HCN), formamide (\ce{NH2CHO}), methyleneimine (\ce{CH2NH}), methylamine (\ce{CH3NH2}), cyanamide (\ce{NH2CN}), and aminoacetonitrile (\ce{NH2CH2CN}) was extremely important because these prebiotic molecules are essential for the production of amino acids in the interstellar medium (ISM) \citep{her09,gar13,bo19, man22a, man22b, man23a}. In the ISM, the hot molecular cores are one of the earlier stages of high-mass star-formation regions, and they play a crucial role in enhancing the chemical complexity in the ISM \citep{shi21}. The hot molecular cores are ideal candidates to study the complex prebiotic molecules because they contain a compact and small source size ($<$0.1 pc), a warm environment ($>$100 K), and a high gas density ($n_{\ce{H2}}$$>$10$^{6}$ cm$^{-3}$) that promotes molecular evolution by thermal hopping on dust grains \citep{van98, wi14}. The chemistry of the hot cores is characterised by the sublimation of ice mantles, which accumulated during the star-formation activities \citep{shi21}. The gaseous molecules and atoms are frozen onto the dust grains in prestellar cores and cold molecular clouds. As the temperature of the dust rises due to star-formation activities, the chemical interactions between heavy species become active on the grain surfaces, resulting in the formation of more complex organic molecules \citep{gar06, shi21}. In addition, sublimated molecules such as methanol (\ce{CH3OH}) and ammonia (\ce{NH3}) are also subjected to further gas-phase reactions \citep{nom04, taq16}. As a result, the warm, dense, and chemically abundant gas around the protostars develops into one of the strongest molecular line emitters, known as the hot molecular cores. The hot molecular cores are crucial objects for astrochemical studies because a multiplicity of simple and complex organic molecules are frequently found in the hot cores \citep{her09}. The period of the hot molecular cores is thought to last about 10$^{5}$ years to 10$^{6}$ years \citep{van98,vi04,gar06, gar08, gar13}.

\begin{table*}
	\centering
	\caption{Observation summary of G358.93--0.03.}
		\begin{adjustbox}{width=1.0\textwidth}
	\begin{tabular}{ccccccccccccccccc}
		\hline 
		Band&Date of observation& Start time&End time &Number  &Frequency range&Spectral resolution & Angular resolution\\
		&(yyyy mm dd)       &(hh:mm) &  (hh:mm)      &  of antennas                  & (GHz)	    & (kHz)              &($\prime\prime$)\\
		\hline
		7&2019 Nov 11&20:50&21:20&47&290.51--292.39&976.56&0.42\\
		&           &     &     &  &292.49--294.37&976.56&\\
		&           &     &     &  &302.62--304.49&976.56&\\
		&           &     &     &  &304.14--306.01&976.56&\\
		
		\hline
	\end{tabular}	
	\label{tab:obs}
		\end{adjustbox}
\end{table*}

In recent years, several molecular species that are recognised as precursors to biologically relevant molecules have been identified in the ISM at millimeter and sub-millimeter wavelengths. The search for amino acids and their possible precursors in the hot molecular cores is one of the most important topics in modern astrochemistry because, earlier, several authors claimed the hot molecular core is an ideal candidate for searching for amino acids and their possible precursors \citep{hal13, gar13, ohi19, suz19, bo19}. Previously, all surveys of the simplest amino acid glycine (\ce{NH2CH2COOH}) towards the hot molecular cores (such as Sgr-B2, Orion KL, etc.) failed because the predicted emission lines of \ce{NH2CH2COOH} were blended with other nearby molecular transitions \citep{jon07, cun07}. Earlier, \cite{kua03} claimed to have made the first \ce{NH2CH2COOH} detections towards the Orion KL, Sgr B2 (N-LMH), and W51 e1/e2, but multiple follow-up observations revealed that these \ce{NH2CH2COOH} detections were false \citep{jon07}. The Rosetta Orbiter Spectrometer for Ion and Neutral Analysis (ROSINA) mass spectrometer has identified volatile \ce{NH2CH2COOH} from the coma of the comet 67P/Churyumov-Gerasimenko \citep{alt16}. Earlier, all investigations of the \ce{NH2CH2COOH} towards the hot molecular cores failed due to the low-sensitivity telescopes, but recently, high-resolution ALMA has been designed specifically to detect the amino acids from the ISM. Earlier, \citet{jim14} computed a 1D spherically symmetric radiative transfer model, and they claimed that there are high chances of detecting the emission lines of \ce{NH2CH2COOH} from the low-mass pre-stellar core L1544 using ALMA bands 3 and 4. Previously, \citet{gar13} predicted that the emission lines of \ce{NH2CH2COOH} would be detectable using ALMA with one hour of on-source integration time towards the NGC 6334 IRS1.

The complex prebiotic molecule \ce{CH3NH2} contains the simplest primary amine (R--\ce{NH2}) functional group and that molecule has been suggested as a possible precursor to the simplest amino acid, \ce{NH2CH2COOH} \citep{kim11, thu11, gar13}. The interstellar molecule \ce{CH3NH2} was first detected in the high-mass star-formation region Sgr B2, which is located in the galactic centre \citep{kai74}. Later, \citet{hal13} detected individual 170 transition lines of \ce{CH3NH2} towards the Sgr B2 (N) with a fractional abundance of 1.7$\times$10$^{-9}$. Earlier, the tentative absorption lines of \ce{CH3NH2} were detected from the quasar PKS 1830--211 \citep{mu11}. Earlier, \cite{lig15} made a spectral line survey towards the nine hot molecular cores using the James Clerk Maxwell Telescope (JCMT), but they could not identify that molecule within the 3$\sigma$ significance. The evidence of \ce{CH3NH2} was also found from the coma of the comet 67P/Churyumov-Gerasimenko \citep{alt16}. The abundance of \ce{CH3NH2} with respect to \ce{NH2CH2COOH} in the comet 67P/Churyumov-Gerasimenko is 1.0$\pm$0.5 \citep{alt16}. The column density of \ce{CH3NH2} towards 67P/Churyumov-Gerasimenko is 1.18$\times$10$^{17}$ cm$^{-2}$ \citep{ma18}. The emission lines of \ce{CH3NH2} were also detected towards the hot molecular core region NGC 6334I MM1--MM3 using ALMA bands 7 and 10 \citep{bo19}. The evidence of the emission lines of \ce{CH3NH2} was also found towards the hot molecular core object G10.47+0.03, with an estimated fractional abundance of 1.5$\times$10$^{-8}$ \citep{ohi19}. Recently, \citet{pag17} presented the tentative detection of the rotational emission lines of \ce{CH3NH2} towards the Orion KL using ALMA in the frequency range of 219.442--220.127 GHz. Previously, \citet{lig18} attempted to use the ALMA to search for \ce{CH3NH2} emission lines towards the low mass protostar IRAS 16293-2422 B, but they were unable to detect this molecule. Recently, \citet{naz22} made a spectral line survey of \ce{CH3NH2} and other nitrogen-bearing molecules towards the thirty-seven high-mass sources using ALMA band 6.

\begin{figure*}
	\centering
	\includegraphics[width=1.0\textwidth]{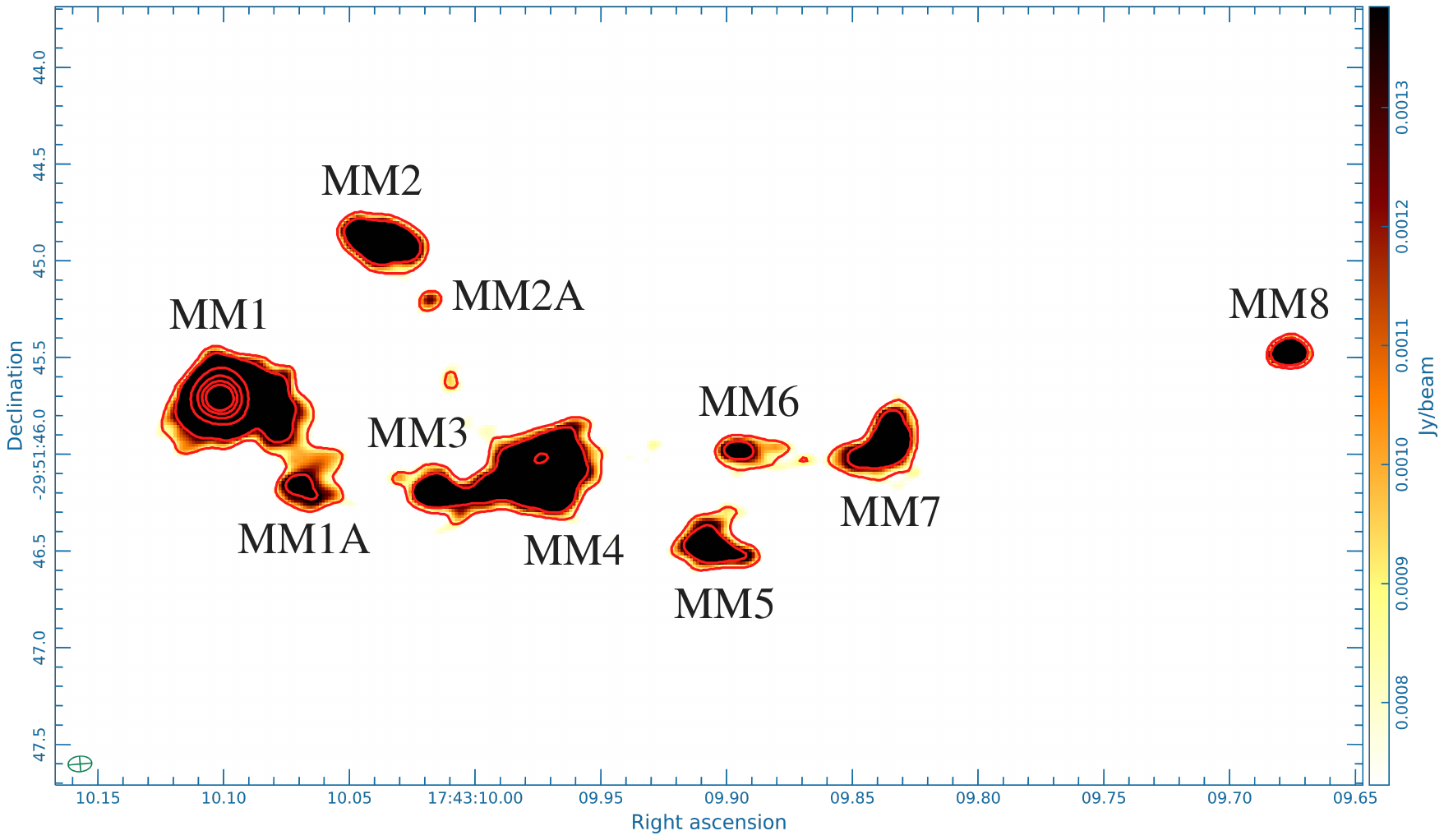}
	\caption{Continuum emission image of the high-mass star-formation region G358.93--0.03 at wavelength 1.02 mm (291.318 GHz). In the continuum emission image, MM1 to MM8 are the continuum sources situated in the massive star formation region G358.93--0.03. The synthesised beam (green circle) size of the continuum emission image was 0.427$^{\prime\prime}$$\times$0.376$^{\prime\prime}$ with a position angle -78.467$^{\circ}$.}
	\label{fig:continuum}
\end{figure*}

The massive star-formation region G358.93--0.03 (RA: 17$^{h}$43$^{m}$10$^{s}$.02, Dec: -29$^{\circ}$51$^{\prime}$45.8$^{\prime\prime}$) is located at a distance of 6.75$\,{}\,^{\,+0.37}_{\,-0.68}$\, kpc \citep{re14,bro19}. The high-mass star-formation region G358.93--0.03 contains eight sub-millimeter continuum sources, which are designated MM1--MM8 in order of decreasing right ascension (R.A) \citep{bro19}. The luminosity of G358.93--0.03 is $\sim$7.7$\times$10$^{3}$ \textup{L}$_{\odot}$ and the total gas mass is 167$\pm$12\textup{M}$_{\odot}$ \citep{bro19}. Among the eight continuum sources, MM1 and MM3 (hereafter, G358.93--0.03 MM1 and G358.93--0.03 MM3) are the brightest continuum sources, which host line-rich hot molecular cores \citep{bro19, bay22}. Earlier, \cite{bro19} and \cite{ste21} discussed the physical properties of hot molecular cores G358.93--0.03 MM1 and G358.93--0.03 MM3. Recently, \citet{bro19} discovered fourteen methanol (\ce{CH3OH}) maser lines from G358.93--0.03 MM1, and they also detected the emission lines of methyl cyanide (\ce{CH3CN}) from G358.93--0.03 MM1 and G358.93--0.03 MM3 using ALMA, but the authors do not estimate the abundance of \ce{CH3CN}. \citet{bro19} also do not describe the formation mechanism of \ce{CH3OH} and \ce{CH3CN} towards the G358.93--0.03 MM1. In the hot molecular cores, \ce{CH3OH} is a highly abundant molecule, and it is formed on the grain surface via the hydrogenation process from CO (CO$\stackrel{\ce{H2}}{\longrightarrow}$\ce{H2CO}$\stackrel{\rm H_{2}}\longrightarrow$\ce{CH3OH}) \citep{gar13, suz16}. Similarly, the \ce{CH3CN} molecule was formed in the hot molecular cores via the radiative association of \ce{CH3}$^{+}$ with HCN or CN with \ce{CH3} \citep{cha92,mil97}. Recently, \citet{chen20} detected the maser emission lines of deuterated water (HDO), isocyanic acid (HNCO), and $^{13}$\ce{CH3OH} towards the G358.93--0.03 using the TMRT and VLA radio telescopes. Recently, the rotational emission lines of the possible urea (\ce{NH2CONH2}) precursor molecule cyanamide (\ce{NH2CN}) and simplest sugar-like molecule glycolaldehyde (\ce{CH2OHCHO}) were also detected towards the G358.93--0.03 MM1 using ALMA \citep{man23a, man23b}. The detection of \ce{CH3CN} and \ce{NH2CN} indicates that other nitrogen-bearing molecules such as HCN, \ce{C2H5CN}, \ce{C2H3CN}, and \ce{C3H7CN} would be observable towards the G358.93--0.03 MM1. We first attempted to search for possible glycine precursor molecules towards the G358.93--0.03 MM1 using ALMA band 7.

In this article, we present the first detection of a possible \ce{NH2CH2COOH} precursor molecule, \ce{CH3NH2}, towards the hot molecular core region G358.93--0.03 MM1 using ALMA. We applied the local thermodynamic equilibrium (LTE) model to estimate the column density and gas temperature of \ce{CH3NH2}. We also discussed the possible formation mechanism of \ce{CH3NH2} towards the G358.93--0.03 MM1. The ALMA observations and data reduction were presented in Section~\ref{obs}. The result of the detection of emission lines of \ce{CH3NH2} was shown in Section~\ref{res}. The discussion and conclusion of the detection of \ce{CH3NH2} towards the G358.93--0.03 MM1 were shown in Section~\ref{dis} and \ref{conclu}.

\begin{figure*}
	\centering
	\includegraphics[width=1.0\textwidth]{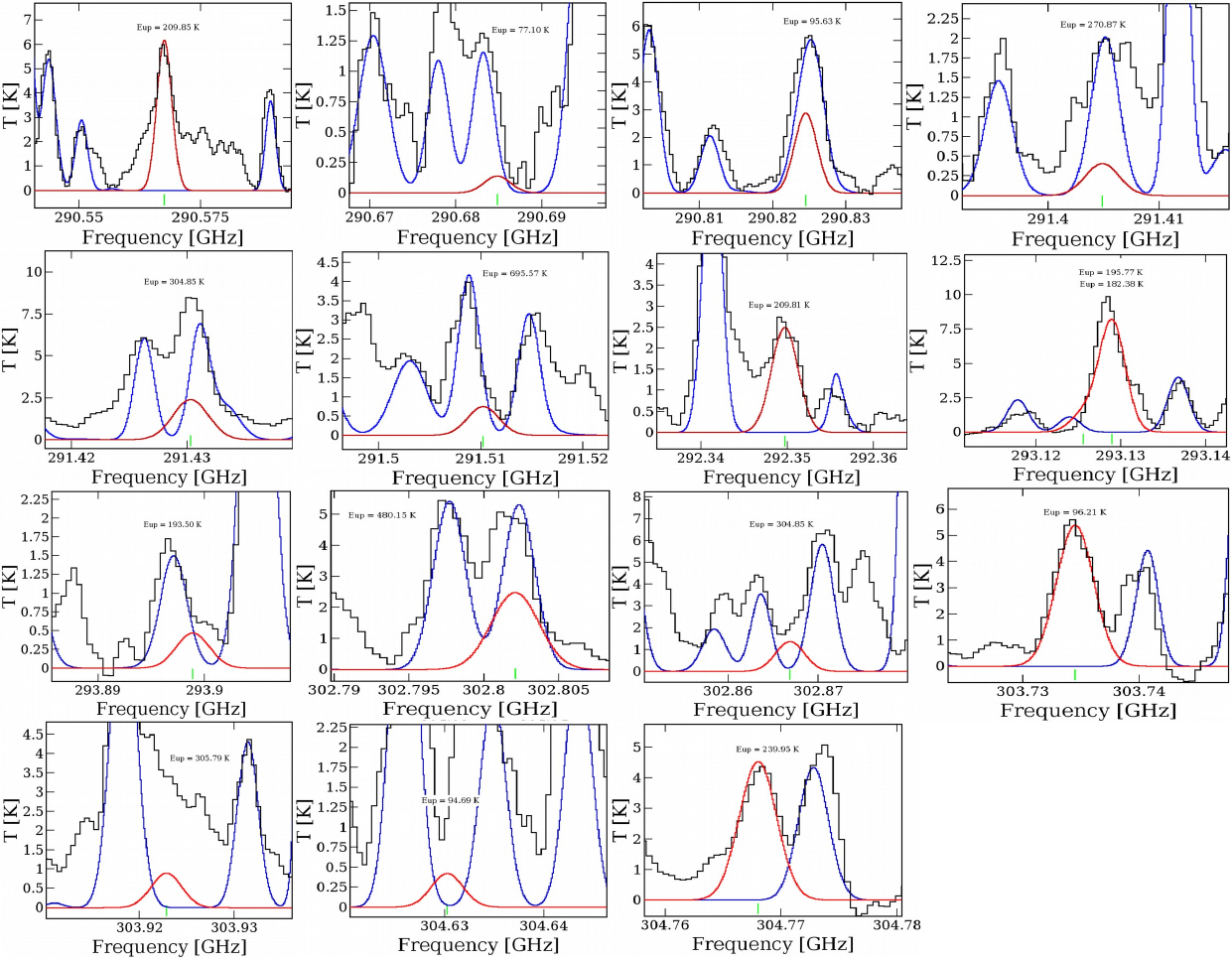}
	\caption{Rotational emission lines of \ce{CH3NH2} towards G358.93--0.03 MM1 in the frequency ranges of 290.51--292.39 GHz, 292.49--294.37 GHz, 302.62--304.49 GHz, and 304.14--306.01 GHz. The black lines indicated the millimeter wavelength molecular spectra of G358.93--0.03 MM1, and the red lines represent the LTE model spectra of \ce{CH3NH2}. The blue lines indicate the LTE spectra of other 250 molecular transitions towards G358.93--0.03 MM1. The radial velocity of the spectra was --16.50 km s$^{-1}$.}
	\label{fig:ltespec}
\end{figure*}

\section{Observations and data reductions}
\label{obs}
The high-mass star-formation region G358.93--0.03 was observed to study the massive protostellar accretion outburst using the Atacama Large Millimeter/Submillimeter Array (ALMA) band 7 (covering the frequency range 290.51--306.01 GHz) receiver (PI: Crystal Brogan). The observation of G358.93--0.03 was performed on November 11, 2019, with phase center ($\alpha,\delta$)$_{\rm J2000}$ = 17:43:10.000, --29:51:46.000 with on-source integration time 756.0 sec. During the observation, a total of 47 antennas were used. For ALMA band 7, the minimum baseline was 14 m and the maximum baseline was 2517 m. During the observation of G358.93--0.03, J1550+0527 was taken as a flux calibrator and bandpass calibrator, and J1744--3116 was taken as a phase calibrator. The observation summary is presented in Table~\ref{tab:obs}.

We used the Common Astronomy Software Application ({\tt CASA 5.4.1}) for data reduction and imaging with the ALMA data reduction pipeline \citep{mc07}. We used the Perley-Butler 2017 flux calibrator model for flux calibration with flux calibrator J1550+0527 for each baseline using the task {\tt SETJY} \citep{pal17}. We used the {\tt CASA} pipeline with tasks {\tt hifa\_bandpassflag} and {\tt hifa\_flagdata} to construct the flux and bandpass calibration after flagging the bad antenna data. We utilised the task {\tt MSTRANSFORM} to divide the target data set into another data set with all available rest frequencies after the initial data calibration. We create the continuum emission image of G358.93--0.03 using the task {TCLEAN} with line-free channels. The continuum emission image of G358.93--0.03 at frequency 291.318 GHz is shown in Figure~\ref{fig:continuum}. In the continuum emission image of G358.93--0.03, we observed all submillimeter continuum sources. Recently, \cite{man23a} and \cite{man23b} discussed in detail the continuum emission from G358.93--0.03 using ALMA. We used the task {\tt UVCONTSUB} in the UV plane of split calibrated data for the continuum subtraction operation. For each rest frequency, we used task {\tt TCLEAN} with a Briggs weighting robust value of 0.5 to construct spectral images of G358.93--0.03. For the correction of the primary beam pattern in the continuum and spectral images, we used the CASA task {\tt IMPBCOR}.

\section{Result}
\label{res}

\begin{table*}
	\centering
	\scriptsize 
	\caption{Summary of the LTE fitted line parameters of the \ce{CH3NH2} towards G358.93--0.03 MM1.}
	\begin{adjustbox}{width=1.0\textwidth}
	\begin{tabular}{ccccccccccccccccc}
		\hline 
		Observed frequency &Transition & $E_{u}$ & $A_{ij}$ &g$_{up}$&$\mu^{2}S$$^{\dagger}$ &FWHM &Optical depth&Remark\\
		
		(GHz) &(${\rm J^{'}_{K_a^{'}\Gamma^{'}}}$--${\rm J^{''}_{K_a^{''}\Gamma^{''}}}$) &(K)&(s$^{-1}$) & &(Debye$^{2}$) &(km s$^{-1}$) &($\tau$)& \\
		\hline
		290.567&13(2)$E_{1+1}$--13(1)$E_{1-1}$&209.85&6.10$\times$10$^{-5}$&324&69.19&3.50$\pm$0.56&5.81$\times$10$^{-2}$&Non blended\\
		
		290.684&8(0)$E_{2+1}$--7(1)$E_{2-1}$&77.10&4.07$\times$10$^{-6}$&68&0.96&3.50$\pm$0.98&1.26$\times$10$^{-3}$&Blended with c-\ce{C3H2}\\
		
		290.824&9(0)$A_{2}$--8(1)$A_{1}$&95.63&8.18$\times$10$^{-5}$&76&21.71&3.50$\pm$0.39&2.67$\times$10$^{-2}$&Blended with \ce{CH3OCHO}\\
		
		291.404&15(2)$E_{1-1}$--15(1)$E_{1-1}$&270.87&4.18$\times$10$^{-6}$&372&5.40&3.50$\pm$0.41&3.71$\times$10$^{-3}$&Blended with \ce{C2H5CN}\\
		
		291.430&16(2)$E_{1-1}$--15(3)$E_{1-1}$&304.85&2.63$\times$10$^{-5}$&396&36.09&3.50$\pm$0.59&2.21$\times$10$^{-2}$&Blended with \ce{C2H4O}\\	
		
		291.510&20(8)$E_{1-1}$--21(7)$E_{1-1}$&695.57&2.40$\times$10$^{-5}$&492&40.97&3.50$\pm$0.62&6.84$\times$10$^{-3}$&Blended with \ce{C3H8}\\
		
		292.349&13(2)$E_{2-1}$--13(1)$E_{2-1}$&209.81&7.30$\times$10$^{-5}$&108&27.11&3.50$\pm$0.97&2.29$\times$10$^{-2}$&Non blended\\
		
		293.125&13(1)$E_{2-1}$--12(2)$E_{2+1}$&195.77&3.56$\times$10$^{-5}$&108&13.12&3.50$\pm$0.65&6.91$\times$10$^{-3}$&Partially Blended with \ce{H2}$^{13}$CO\\
		
		293.128&12(2)$B_{1}$--12(1)$B_{2}$&182.38&8.00$\times$10$^{-5}$&300&81.85&3.50$\pm$0.39&4.50$\times$10$^{-2}$&Non blended \\
		
		293.898&9(5)$E_{2+1}$--10(4)$E_{2+1}$&193.50&1.84$\times$10$^{-5}$& 76&4.72&3.50$\pm$0.47&4.48$\times$10$^{-3}$&Blended with OS$^{17}$O\\
		
		~302.802$^{*}$&16(7)$B_{1}$--17(6)$B_{2}$&480.15&2.50$\times$10$^{-5}$&396&30.58&3.50$\pm$0.37&1.15$\times$10$^{-2}$&Blended with \ce{CH3CHO}\\
		
		302.866&16(2)$E_{1-1}$--15(3)$E_{1+1}$&304.85&1.53$\times$10$^{-5}$&396&18.69&3.50$\pm$0.45&1.26$\times$10$^{-2}$&Blended with \ce{C2H5CN}\\
		
		303.734&9(0)$E_{2+1}$--8(1)$E_{2+1}$&96.21&8.94$\times$10$^{-5}$&76&20.82&3.50$\pm$0.79&2.82$\times$10$^{-2}$&Non blended\\
		
		303.922&12(6)$E_{2-1}$--13(5)$E_{2-1}$&305.79&2.22$\times$10$^{-5}$&100&6.78&3.50$\pm$0.71&4.58$\times$10$^{-3}$&Blended with \ce{CH2CHCN}\\

		304.630&5(4)$E_{2-1}$--6(3)$E_{2-1}$&94.69&1.18$\times$10$^{-5}$&44&1.57&3.50$\pm$0.57&2.15$\times$10$^{-3}$&Blended with \ce{CH2DOH}\\
		
		304.768&14(2)$A_{1}$--14(1)$A_{2}$&239.95&7.94$\times$10$^{-5}$&116&27.96&3.50$\pm$0.25&2.36$\times$10$^{-2}$&Non blended\\
		\hline
	\end{tabular}	
		\end{adjustbox}
	\label{tab:MOLECULAR DATA}\\
	{{*}}--The transition of \ce{CH3NH2} contain double with frequency difference  $\leq$100 kHz. The second transition is not shown.\\
	$\dagger$--The line intensity $\mu^{2}S$ is defined by the product of the square of the dipole moment $\mu^{2}$ and transition line strength $S$. The values of $\mu^{2}S$ of detected transitions of \ce{CH3NH2} are taken from the JPL molecular database. \\
\end{table*}

\subsection{Line emission towards G358.93--0.03}
From the spectral images of G358.93--0.03, we observed that only the spectra of hot molecular cores G358.93--0.03 MM1 and G358.93--0.03 MM3 show any line emission. We cannot see any line emission from other sources in G358.93--0.03. The synthesized beam sizes of the spectral images of G358.93--0.03 at frequency ranges of 290.51--292.39 GHz, 292.49--294.37 GHz, 302.62--304.49 GHz, and 304.14--306.01 GHz are 0.425$^{\prime\prime}\times$0.369$^{\prime\prime}$, 0.427$^{\prime\prime}\times$0.376$^{\prime\prime}$, 0.413$^{\prime\prime}\times$0.364$^{\prime\prime}$, and 0.410$^{\prime\prime}\times$0.358$^{\prime\prime}$, respectively. We generate the molecular spectra from G358.93--0.03 MM1 and G358.93--0.03 MM3 by creating a 0.91$^{\prime\prime}$ diameter circular region, which is larger than the line emitting regions of G358.93--0.03 MM1 and G358.93--0.03 MM3. The position of G358.93--0.03 MM1 is RA (J2000) = 17$^{h}$43$^{m}$10$^{s}$.101, Dec (J2000) = --29$^\circ$51$^{\prime}$45$^{\prime\prime}$.693. The position of G358.93--0.03 MM3 is RA (J2000) = 17$^{h}$43$^{m}$10$^{s}$.0144, Dec (J2000) = --29$^\circ$51$^{\prime}$46$^{\prime\prime}$.193. Recently, \citet{man23b} shows the resultant spectra of G358.93--0.03 MM1 and G358.93--0.03 MM3 between the frequency range of 290.51--306.01 GHz in Figure 2. \citet{man23b} shows that the molecular spectra of G358.93--0.03 MM1 are more chemically rich than G358.93--0.03 MM3. From the molecular spectra of G358.93--0.03 MM1, \citet{man23b} observes the inverse P-Cygni profile associated with \ce{CH3OH} lines towards the G358.93--0.03 MM1. This may indicate that the G358.93--0.03 MM1 is undergoing an infall. \citet{man23b} does not observe any inverse P-cygni profile in the spectra of G358.93--0.03 MM3. The systemic velocities ($V_{LSR}$) of G358.93--0.03 MM1 and G358.93--0.03 MM3 are --16.5 km s$^{-1}$ and --18.2 km s$^{-1}$, respectively \citep{bro19}.

\subsubsection{Identification of \ce{CH3NH2} towards G358.93--0.03 MM1}
To identify the rotational emission lines of \ce{CH3NH2} towards the G358.93--0.03 MM1, we used the local thermodynamic equilibrium (LTE) model with the Jet Population Laboratory (JPL) \citep{pic88} spectroscopic molecular database. We have used the LTE-RADEX module in CASSIS for the LTE modelling \citep{vas15}. The LTE assumptions are reasonable in the inner region of the hot molecular core candidate G358.93--0.03 MM1 because the gas density of the warm inner region of the hot core was 2$\times$10$^{7}$ cm$^{-3}$ \citep{ste21}. Additionally, we also fitted the LTE model to the other 250 molecules over the observed molecular spectra of G358.93--0.03 MM1 to understand the blended effect between other molecules and \ce{CH3NH2}, including those molecules detected by \citet{man23a} and \citet{man23b}. The LTE-fitted rotational emission spectra of \ce{CH3NH2} are shown in Figure~\ref{fig:ltespec}. After the LTE analysis, we observed that the sixteen transition lines of \ce{CH3NH2} are present within the observed frequency ranges, out of which five lines of \ce{CH3NH2} are non-blended and clearly detected, while the remaining ones are blended with other lines. All the non-blended emission lines of \ce{CH3NH2} are identified as higher than 5$\sigma$ (confirmed from LTE modelling). The upper-level energies of the identified sixteen transitions of \ce{CH3NH2} vary from 77.10 K to 695.57 K. The upper-level energies of the non-blended transitions of \ce{CH3NH2} vary between 96.21 K and 239.95 K. All the transitions of \ce{CH3NH2}, generated by LTE model (non-blended or blended), are detected in the observable frequency ranges. The best-fit column density of \ce{CH3NH2} was (1.10$\pm$0.31)$\times$10$^{17}$ cm$^{-2}$ with an excitation temperature of 180.8$\pm$25.5 K and a source size of 0.5$^{\prime\prime}$. Our derived column density of \ce{CH3NH2} towards G358.93--0.03 MM1 is similar to the column density of \ce{CH3NH2} in the comet 67P/Churyumov-Gerasimenko. Additionally, we also detected the emission lines of \ce{C2H5CN}, \ce{C2H3CN}, \ce{CH3OCHO}, \ce{CH3SH}, \ce{CH3OH}, \ce{NH2CHO}, \ce{CH3CN}, and \ce{HC3N}, which we discuss in a separate paper.

After the identification of the rotational emission lines of \ce{CH3NH2} using the LTE model, we obtained the quantum numbers ({${\rm J^{'}_{K_a^{'}\Gamma^{'}}}$--${\rm J^{''}_{K_a^{''}\Gamma^{''}}}$}), upper state energy ($E_u$) in K, Einstein coefficients ($A_{ij}$) in s$^{-1}$, line intensity ($\mu^{2}S$) in Debye$^{2}$, full-width half maximum (FWHM) in km s$^{-1}$, and optical depth ($\tau$). The transitions of \ce{CH3NH2} are labelled by the torsion-inversion rotation between the primary amine (R--\ce{NH2}) and the methyl group (R--\ce{CH3}), which was represented as $\Gamma$. The $\Gamma$ states are labelled with the $A_{1}$, $A_{2}$, $B_{1}$, $B_{2}$, $E_{1\pm1}$, and E$_{2\pm1}$ \citep{ily07}. The selection rules of \ce{CH3NH2} are $\Gamma$ = $A_{1}$$\leftrightarrow$$A_{2}$, $B_{1}$$\leftrightarrow$$B_{2}$, $E_{1\pm1}$$\leftrightarrow$$E_{1\pm1}$, and E$_{2\pm1}$$\leftrightarrow$E$_{2\pm1}$. The symmetry levels of \ce{CH3NH2} have nuclear spin-statistical weights of 1 for the $A_{1}$, $A_{2}$, and $E_{2}$ states and 3 for the $B_{1}$ , $B_{2}$, and $E_{1}$ states. The hyperfine structures of \ce{CH3NH2} are not considered in this analysis because the line intensity ($\mu^{2}S$) of the main transitions of \ce{CH3NH2} is higher than hyperfine lines and current spectral resolution is not sufficient to resolve the hyperfine lines of \ce{CH3NH2}. After the LTE analysis, we observed J = 13(2)$E_{1+1}$--13(1)$E_{1-1}$, J = 13(2)$E_{2-1}$--13(1)$E_{2-1}$, J = 12(2)$B_{1}$--12(1)$B_{2}$, J = 9(0)$E_{2+1}$--8(1)$E_{2+1}$, and J = 14(2)$A_{1}$--14(1)$A_{2}$ transition lines of \ce{CH3NH2} are not contaminated with other nearby molecular transitions. From Figure~\ref{fig:ltespec}, it is clear that all LTE-fitted lines of \ce{CH3NH2} are consistent with the observed spectra, but some of them are contaminated by stronger lines from other species. The summary of the detected transitions and spectral line properties of \ce{CH3NH2} was presented in Table~\ref{tab:MOLECULAR DATA}.

To determine the fractional abundance of \ce{CH3NH2}, we used the column density of \ce{CH3NH2} inside the 0.45$^{\prime\prime}$ beam, which we divided by the \ce{H2} column density of G358.93--0.03 MM1. The fractional abundance of \ce{CH3NH2} with respect to \ce{H2} towards the G358.93--0.03 MM1 is (8.80$\pm$2.60)$\times$10$^{-8}$, where the column density of \ce{H2} towards the G358.93--0.03 MM1 is (1.25$\pm$0.11)$\times$10$^{24}$ cm$^{-2}$ \citep{man23a}. The column density ratio of \ce{CH3NH2} and \ce{NH2CN} towards G358.93--0.03 MM1 is (1.86$\pm$0.95)$\times$10$^{2}$, where the column density of \ce{NH2CN} towards G358.93--0.03 MM1 is (5.9$\pm$2.5)$\times$10$^{14}$ cm$^{-2}$\citep{man23a}. We estimate the column density ratio of \ce{CH3NH2} and \ce{NH2CN} because \ce{CH3NH2} acts as a possible precursor of \ce{NH2CN} in the gas-phase, and both molecules share \ce{NH2} as a common precursor \citep{ram23}. Earlier, \cite{ohi19} found that the abundance of \ce{CH3NH2} towards another hot molecular core, G10.47+0.03, was (1.5$\pm$1.1)$\times$10$^{-8}$, which is nearly similar to our derived abundance of \ce{CH3NH2} towards G358.93--0.03 MM1. This may indicate that the chemical formation route(s) of \ce{CH3NH2} towards the G358.93--0.03 MM1 are similar to those in G10.47+0.03.

\begin{figure}
	\centering
	\includegraphics[width=0.48\textwidth]{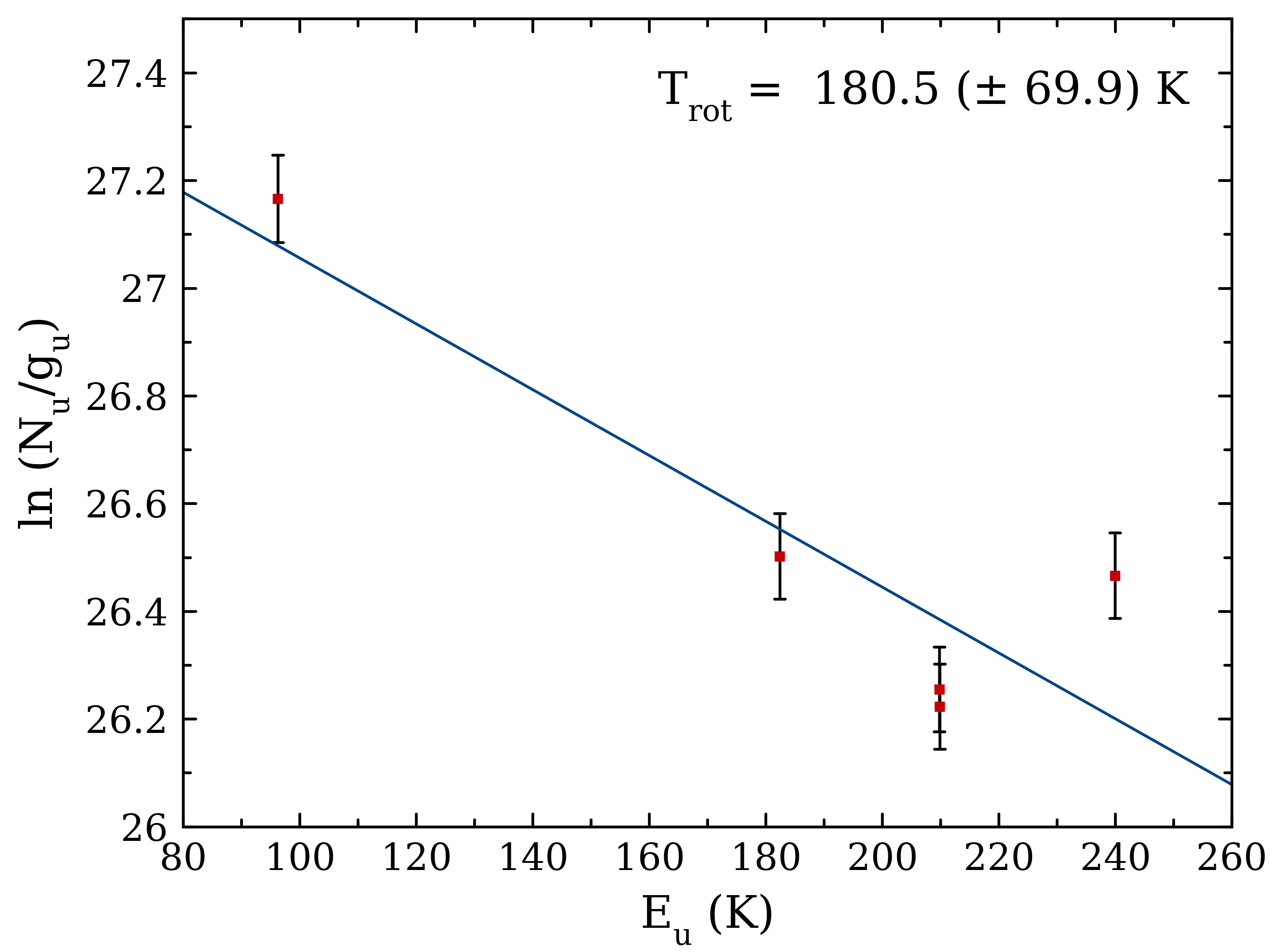}
	\caption{Rotational diagram of \ce{CH3NH2} towards hot molecular core region G358.93--0.03 MM1. In the rotational diagram, the red blocks presented the statistical data points, and the solid blue line indicated the fitted straight line that estimated the rotational temperature of \ce{CH3NH2}.}
	\label{fig:rotd}
\end{figure} 

\subsubsection{Rotational diagram analysis: estimation of the rotational temperature of \ce{CH3NH2}}
To verify the estimated temperature from the LTE modelling, we have created the rotational diagram based on the five non-blended transitions of \ce{CH3NH2}. Initially, we assumed that the identified \ce{CH3NH2} emission lines were optically thin and that they were populated under local thermal equilibrium (LTE) conditions. For optically thin molecular emission lines, the equation of column density can be written as \citep{gold99},
\begin{equation}
{N_u^{thin}}=\frac{3{g_u}k_B\int{T_{mb}dV}}{8\pi^{3}\nu S\mu^{2}}
\end{equation}
where $\int T_{mb}dV$ indicated the integrated intensity, $k_B$ is the Boltzmann constant, $\mu$ indicated the electric dipole moment, $g_u$ is the degeneracy of the upper state, $S$ indicated the strength of the transition lines, and $\nu$ is the rest frequency. The total column density of \ce{CH3NH2} under LTE conditions can be written as,

\begin{equation}
\frac{N_u^{thin}}{g_u} = \frac{N_{total}}{Q(T_{rot})}\exp(-E_u/k_BT_{rot})
\end{equation}
where $T_{rot}$ is the rotational temperature of \ce{CH3NH2}, $E_u$ is the upper state energy of \ce{CH3NH2}, and ${Q(T_{rot})}$ defines the partition function at the extracted rotational temperature. The rotational partition function of \ce{CH3NH2} at 75 K is 15560, at 150 K is 44047, and at 300 K is 124715. Equation 2 can be rearranged as,
\begin{equation}
ln\left(\frac{N_u^{thin}}{g_u}\right) = ln(N)-ln(Q)-\left(\frac{E_u}{k_BT_{rot}}\right)
\end{equation}

The equation 3 indicated a linear relationship between $\ln(N_{u}/g_{u}$) and the upper state energy ($E_{u}$) of \ce{CH3NH2}. The value $\ln(N_{u}/g_{u}$) was calculated using the equation 1. Equation 3 indicates that the spectral parameters with respect to different transition lines of \ce{CH3NH2} should be fitted with a straight line whose slope is inversely proportional to rotational temperature ($T_{rot}$). For the rotational diagram, we estimated the spectral line parameters of \ce{CH3NH2} after fitting a Gaussian model over the non-blended observed spectra of \ce{CH3NH2} using the line analysis module in CASSIS. The resultant rotational diagram of \ce{CH3NH2} is shown in Figure~\ref{fig:rotd}. In the rotational diagram, the vertical black error bars indicate the absolute uncertainty of $\ln(N_{u}/g_{u}$), which was estimated from the $\int T_{mb}dV$. The rotational diagram yields the rotational temperature of 180.5$\pm$69.9 K. This result agrees reasonably well with the value of 180.8$\pm$25.5 K, which is derived by the LTE spectral modelling.

\begin{figure*}
	\centering
	\includegraphics[width=0.95\textwidth]{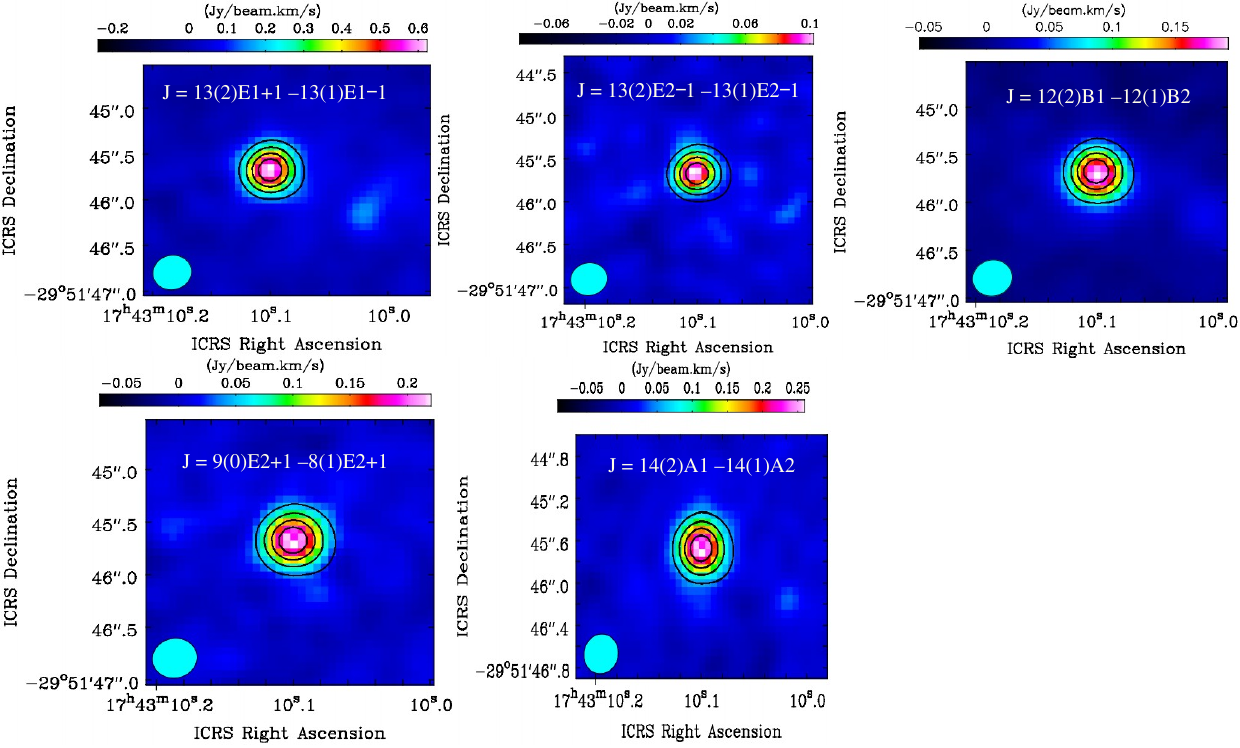}
	\caption{Integrated emission maps of non-blended transitions of \ce{CH3NH2} towards G358.93--0.03 MM1. The integrated emission maps of \ce{CH3NH2} are overlaid with the 1.02 mm (291.318 GHz) continuum emission map (black contour). The contour levels are at 20\%, 40\%, 60\%, and 80\% of the peak flux. The cyan circle represents the synthesised beam of the integrated emission maps.}
	\label{fig:emi}
\end{figure*}

\subsubsection{Searching for \ce{CH3NH2} towards G358.93--0.03 MM3}
After the successful detection of the \ce{CH3NH2} towards the G358.93--0.03 MM1, we also searched for emission lines of \ce{CH3NH2} towards the G358.93--0.03 MM3 using the LTE-modelled spectra, which yielded no detection. The derived upper limit column density of \ce{CH3NH2} towards the G358.93--0.03 MM3 is $\leq$(5.21$\pm$1.23)$\times$10$^{15}$ cm$^{-2}$. The upper limit of the fractional abundance of \ce{CH3NH2} towards the G358.93--0.03 MM3 is $\leq$(1.48$\pm$0.46)$\times$10$^{-8}$. The column density of \ce{H2} towards the G358.93--0.03 MM3 was (3.51$\pm$0.7)$\times$10$^{23}$ cm$^{-2}$ \citep{man23b}.

\begin{table}{}
	\centering
	\caption{Estimated emitting regions of non-blended transitions of \ce{CH3NH2} towards G358.93--0.03 MM1.}
	\begin{tabular}{cccccccccccc}
		\hline
		Frequency&Transition& E$_{up}$&Emitting region\\
		
		GHz&(${\rm J^{'}_{K_a^{'}K_c^{'}}}$--${\rm J^{''}_{K_a^{''}K_c^{''}}}$)&(K)&($^{\prime\prime}$)\\
		\hline
		290.567&13(2)$E_{1+1}$--13(1)$E_{1-1}$&209.85&0.423\\
		292.349&13(2)$E_{2-1}$--13(1)$E_{2-1}$&209.81&0.427\\
		293.128&12(2)$B_{1}$--12(1)$B_{2}$&182.38&0.425\\
		303.734&9(0)$E_{2+1}$--8(1)$E_{2+1}$&96.21&0.430\\
		304.768&14(2)$A_{1}$--14(1)$A_{2}$&239.95&0.427\\			
		\hline
	\end{tabular}	
	
	
	\label{tab:emittingregion}
\end{table}	

\subsection{Spatial distribution of \ce{CH3NH2} towards G358.93--0.03 MM1}
After the identification of the rotational emission lines of \ce{CH3NH2} from G358.93--0.03 MM1, we created the integrated emission maps of \ce{CH3NH2} using the CASA task {\tt IMMOMENTS}. The integrated emission maps of \ce{CH3NH2} towards the G358.93--0.03 MM1 were presented in Figure~\ref{fig:emi}. The integrated emission maps of \ce{CH3NH2} towards the G358.93--0.03 MM1 were constructed by integrating the spectral images in the velocity ranges where the emission lines of \ce{CH3NH2} were identified. We created the integrated emission maps only for the five non-blended transitions of \ce{CH3NH2} towards the G358.93--0.03 MM1. The resultant integrated emission maps of \ce{CH3NH2} were overlaid with a 1.02 mm continuum emission image of G358.93--0.03. We noticed that the integrated emission maps of \ce{CH3NH2} exhibit a peak at the position of the continuum. The integrated emission maps indicate that the emission lines of \ce{CH3NH2} originate from the high gas density, the warm inner region of G358.93--0.03 MM1. After the extraction of the integrated emission maps, we estimated the emitting regions of \ce{CH3NH2} towards the G358.93--0.03 MM1 by fitting the 2D Gaussian over the integrated emission map of \ce{CH3NH2} using CASA task {\tt IMFIT}. The deconvolved beam size of the emitting region was estimated by the following equation,\\
\begin{equation}
\theta_{S}=\sqrt{\theta^2_{50}-\theta^2_{beam}}
\end{equation}
where $\theta_{50} = 2\sqrt{A/\pi}$ indicated the diameter of the circle whose area ($A$) was enclosing the 50\% line peak, and $\theta_{beam}$ was the half-power width of the synthesised beam \citep{riv17}. The estimated emitting regions of \ce{CH3NH2} are presented in Table~\ref{tab:emittingregion}. The emitting regions of \ce{CH3NH2} vary between 0.423$^{\prime\prime}$--0.430$^{\prime\prime}$. The synthesised ALMA beams are diluted by a factor of $\sim$0.63 to 0.65. We noticed that the estimated emitting regions of \ce{CH3NH2} are comparable to or slightly greater than the synthesised beam sizes of the integrated emission maps. This demonstrates that the detected \ce{CH3NH2} transition lines were not spatially resolved or were only marginally resolved towards the G358.93--0.03 MM1. As a result, concluding the morphology of the spatial distribution of \ce{CH3NH2} towards the G358.93--0.03 MM1 is impossible. Higher spatial and angular resolution observations are required to understand the spatial distribution of \ce{CH3NH2} towards the G358.93--0.03 MM1.

\section{Discussion}
\label{dis}

\subsection{\ce{CH3NH2} towards G358.93--0.03 MM1 and comparision with other sources}
We present the first confirmed detection of \ce{CH3NH2} towards G358.93--0.03 MM1 with an estimated fractional abundance of (8.80$\pm$2.60)$\times$10$^{-8}$. The estimated excitation temperature of \ce{CH3NH2} is 180.8$\pm$25.5 K, which indicates the emission lines of \ce{CH3NH2} arise from the warm inner region of the G358.93--0.03 MM1 because the temperature of the hot core is above 100 K \citep{van98}. Earlier, \citet{ohi19} detected the rotational emission lines of \ce{CH3NH2} from another hot molecular core object, G10.47+0.03, with a spectral width of 3.1--6.9 km s$^{-1}$. \citet{ohi19} also does not consider the blended effect of \ce{CH3NH2} with other molecular transitions towards the G10.47+0.03. \cite{ohi19} estimated column density of \ce{CH3NH2} towards the G10.47+0.03 was (1.0$\pm$0.7)$\times$10$^{15}$ cm$^{-2}$ with rotational temperature 46$\pm$21 K. The fractional abundance of \ce{CH3NH2} towards the G10.47+0.03 was (1.5$\pm$1.1)$\times$10$^{-8}$ \citep{ohi19}. The measurements of rotational temperature in \cite{ohi19}, is probably erroneous due to the limited resolution of the Nobeyama 45 m single dish radio telescope. Later, \cite{suz19} also detected the rotational emission lines of \ce{CH3NH2} towards the G10.47+0.03 using ALMA with column density 9.4$\times$10$^{17}$ cm$^{-2}$ and rotational temperature 166 K. Aside from G10.47+0.03, \ce{CH3NH2} emission lines were also detected from other hot molecular core candidates, Sgr B2 and Orion KL \citep{hal13, pag17}. The emission lines of \ce{CH3NH2} also detected towards galactic center cloud G+0.693--0.027 \citep{ze18}. The fractional abundances of \ce{CH3NH2} towards the Sgr B2, Orion KL, and G+0.693--0.027 are 1.7$\times$10$^{-9}$, 1.0$\times$10$^{-9}$, 2.2$\times$10$^{-8}$ respectively \citep{hal13, pag17, ze18}. The fractional abundance of \ce{CH3NH2} towards G358.93--0.03 MM1 is approximately one order of magnitude higher as compared to Orion KL and Sgr B2. The abundance of \ce{CH3NH2} towards the G358.93--0.03 MM1 is nearly similar to the G10.47+0.03 and G+0.693--0.027. If we compare the abundance of \ce{CH3NH2} with all sources where \ce{CH3NH2} is detected, then we observe that the G358.93--0.03 MM1 is the most highly abundant \ce{CH3NH2} source ever known.

\begin{table*}
	\caption{Comparison between modelled and observed abundances of \ce{CH3NH2}.}
	\label{tab:comp}
	\begin{adjustbox}{width=1.0\textwidth}
		\centering   
		\begin{tabular}{|c|c|c|c|ccc|ccc|ccc|ccc|c|}
			\hline 
			& \multicolumn{3}{c}{Simulated Values$^{\rm a}$} & &\multicolumn{3}{c}{Observed Values} \\
			\hline
			Species & Fast & Medium & Slow & &G358.93--0.03 MM1$^{\rm b}$ & &&G10.47+0.03$^{\rm c}$ & &&Sgr B2$^{\rm d}$ & &&Orion KL$^{\rm e}$ &\\
			\hline
			&$f$(\ce{CH3NH2})&$f$(\ce{CH3NH2})&$f$(\ce{CH3NH2}) &$f$(\ce{CH3NH2})&Gas density &\textup{L}$_{\odot}$ &$f$(\ce{CH3NH2})&Gas density &\textup{L}$_{\odot}$ &$f$(\ce{CH3NH2})&Gas density &\textup{L}$_{\odot}$&$f$(\ce{CH3NH2})&Gas density &\textup{L}$_{\odot}$\\
			\hline
			\ce{CH3NH2} & $8.0\times10^{-8}$& $3.6\times10^{-8}$& $6.8\times10^{-9}$& (8.80$\pm$2.60)$\times$10$^{-8}$&2$\times$10$^{7}$ cm$^{-3}$  &7.7$\times$10$^{3}$  & (1.5$\pm$1.1)$\times$10$^{-8}$&1$\times$10$^{7}$ cm$^{-3}$ &7$\times$10$^{5}$ &$1.70\times10^{-9}$&1.7$\times$10$^{7}$ cm$^{-3}$ &1.8$\times$10$^{6}$ &1.0$\times$10$^{-9}$&1-10$\times$10$^{7}$ cm$^{-3}$&1$\times$10$^{5}$\\
			\hline
		\end{tabular}
	\end{adjustbox}
	Notes: a -- Values taken from Table\,8 of \cite{gar13}; \\b --  this work\\c --Taken from \citet{ohi19}\\~~d -- Taken from \citet{hal13}\\~~e -- Taken from \citet{pag17}.
	
\end{table*}

\subsection{Possible formation mechanism of \ce{CH3NH2} and linking with \ce{NH2CH2COOH}}
Earlier, \citet{gar08} and \citet{gar13} proposed that \ce{CH3NH2} is formed on the grain surface of the hot molecular cores by the reaction between radical \ce{CH3} and radical \ce{NH2}. The radicals \ce{CH3} and \ce{NH2} are created from the photodissociation of primarily \ce{CH4} and \ce{NH3}. Similarly, \citet{kim11} suggested that the \ce{CH3NH2} molecule will be formed in the hot molecular cores during the warm-up phases due to the reactions between \ce{NH2} and \ce{CH3} when exposed to cosmic rays on the grain surface (\ce{NH2}+\ce{CH3}$\longrightarrow$\ce{CH3NH2}). Alternatively, \citet{thu11} proposed that the \ce{CH3NH2} and \ce{CH2NH} molecules can be created towards hot molecular cores due to hydrogenation of HCN on the dust surface (HCN $\stackrel{\rm 2H}{\longrightarrow}$ CH$_2$NH$\stackrel{\rm 2H}{\longrightarrow}$CH$_3$NH$_2$). Earlier, \citet{wal14} claimed that \ce{CH3NH2} is created on grains at 10 K by the atom addition reactions to solid \ce{CH2NH}. That means \ce{CH2NH} acted as a possible precursor of \ce{CH3NH2} in the hot molecular cores. \cite{lig15} shows an alternative formation route of \ce{CH2NH} and \ce{CH3NH2} by two different gas-phase reaction pathways, 1. \ce{CH2NH} created between the reaction of CH radical and \ce{NH3} in the gas-phase (CH$^{\bullet}$(g) + \ce{NH3}(g)$\longrightarrow$\ce{CH2NH} + H), 2. \ce{CH3NH2} created between the reaction of \ce{CH3} radicals and \ce{NH3} in the gas-phase (\ce{CH3}$^{\bullet}$(g) + \ce{NH3}(g)$\longrightarrow$\ce{CH3NH2} + H). Additional chemical modelling was required to establish whether these gas-phase reactions can effectively reproduce the observed abundance of \ce{CH3NH2}.

The simplest amino acid \ce{NH2CH2COOH} can be created in the solid phase of hot molecular cores when \ce{CH3NH2} thermally reacts with \ce{CO2} under vacuum UV irradiation (\ce{CH3NH2}$\stackrel{\rm CO_{2}}{\longrightarrow}$\ce{NH2CH2COOH}) \citep{hol05}. Alternatively, \citet{gar13} and \citet{suz17} confirmed another efficient chemical pathway for the formation of \ce{NH2CH2COOH} in the hot molecular cores. They claimed the gas-phase reaction between the radical \ce{CH2NH2} and radical HOCO can create \ce{NH2CH2COOH} towards hot molecular cores (\ce{CH2NH2} + HOCO$\longrightarrow$\ce{NH2CH2COOH}). The \ce{CH2NH2} radical can be created on the grain surface of hot molecular cores by the photodissociation of \ce{CH3NH2} (\ce{CH3NH2} + h$\nu$ $\longrightarrow$\ce{CH2NH2} + H) \citep{ohi19}. Similarly, \citet{ohi19} claimed that the radical HOCO would be created when hydrogen atoms react with \ce{CO2} in the grain surface of hot molecular cores (\ce{CO2} + H $\longrightarrow$ HOCO), but we observe \cite{gar13} claimed that the HOCO radical can be created from formic acid (HCOOH), not from \ce{CO2}. After the detection of highly abundant \ce{CH3NH2} towards the G358.93--0.03 MM1, we claimed that the hot molecular core candidate G358.93--0.03 MM1 is the new glycine survey candidate for the future.

\subsection{Previous chemical modelling of \ce{CH3NH2} towards the hot molecular cores}
To understand the formation mechanism of \ce{CH3NH2} towards the hot molecular core, we used the three-phase (gas, grain, and ice mantle) warm-up chemical modelling abundance of \ce{CH3NH2}, which was done by \citet{gar13}. That chemical modelling of \citet{gar13} is mainly used to understand the formation mechanism of \ce{NH2CH2COOH} and its precursors towards the hot molecular cores. During chemical modelling, \cite{gar13} assumed the isothermal collapse phase after a static warm-up phase. In the first phase (free-fall collapse), the gas density rapidly increases from $n_{H}$ = 3$\times$10$^{3}$ to 10$^{7}$ cm$^{-3}$, and the dust temperature reduces to 8 K from 16 K under the free-fall collapse. In the second phase (warm-up phase), the gas density remains constant at $\sim$10$^{7}$ cm$^{-3}$ and the temperature fluctuates rapidly from 8 K to 400 K over the time scales 5$\times$10$^{4}$ yr (fast warm-up), 2$\times$10$^{5}$ yr (medium warm-up), and 1$\times$10$^{6}$ yr (slow warm-up). In the chemical network used by \citet{gar13}, the recombination of the radical \ce{CH3} and radical \ce{NH2} dominates the production of \ce{CH3NH2} on the grains. In the warm-up phase, \citet{gar13} shows the abundance of \ce{CH3NH2} towards the hot molecular cores was 8.0$\times$10$^{-8}$ for the fast warm-up condition, 3.6$\times$10$^{-8}$ for the medium warm-up condition, and 6.8$\times$10$^{-9}$ for the slow warm-up condition.

\subsection{Comparison with modelled and observed abundance of \ce{CH3NH2}}
To understand the formation pathways of \ce{CH3NH2} towards G358.93--0.03 MM1, we compare our estimated abundance of \ce{CH3NH2} with the three-phase warm-up chemical modelled abundance of \ce{CH3NH2}, which were done by \cite{gar13}. This comparison is physically reasonable because the temperature and gas density of G358.93--0.03 MM1 is $\sim$150 K \citep{chen20} and $\sim$2$\times$10$^{7}$ cm$^{-3}$ \citep{ste21}, respectively. Hence, the three-phase warm-up chemical model based on the timescales of \cite{gar13} is appropriate for explaining the chemical evolution of \ce{CH3NH2} towards the G358.93--0.03 MM1. The comparison between the observed abundance and modelled abundance of \ce{CH3NH2} is estimated based on fast, medium, and slow warm-up models, which are presented in Table~\ref{tab:comp}. We noticed that the fast warm-up model result was nearly similar to the observation values of \ce{CH3NH2} towards the G358.93--0.03 MM1.
This comparison indicates that the \ce{CH3NH2} may form on the grain surface via the reaction between radical \ce{CH3} and radical \ce{NH2} towards the G358.93--0.03 MM1. Furthermore, we also compared the fractional abundance of \ce{CH3NH2} towards the G10.47+0.03, Sgr B2, and Orion KL with the modelled abundance of \ce{CH3NH2}, which is estimated by \cite{gar13} as presented in Table.~\ref{tab:comp}. The three-phase warm-up chemical model of \cite{gar13} is also reasonable towards G10.47+0.03, Sgr B2, and Orion KL because the gas density of those hot molecular cores is 1$\times$10$^{7}$ cm$^{-3}$, 1.7$\times$10$^{7}$ cm$^{-3}$, and 1$\times$10$^{7}$ cm$^{-3}$, respectively \citep{hal13, pag17, ohi19}. We noticed that the medium warm-up model satisfied the abundance of \ce{CH3NH2} towards the G10.47+0.03, but the slow warm-up model satisfied the abundance of \ce{CH3NH2} towards the Sgr B2 and Orion KL. That comparison indicates \ce{CH3NH2} may be created on the grain surface via the reaction between radical \ce{CH3} and radical \ce{NH2} towards the G358.93--0.03 MM1, G10.47+0.03, Sgr B2, and Orion KL with different time scales.

\subsection{Searching of \ce{NH2CH2COOH} towards G358.93--0.03 MM1}  
After the detection of a possible \ce{NH2CH2COOH} precursor molecule \ce{CH3NH2} towards G358.93--0.03 MM1, we also searched the emission lines of the simplest amino acid \ce{NH2CH2COOH} conformers I and II towards the G358.93--0.03 MM1. After spectral analysis using the LTE model spectra of \ce{NH2CH2COOH} conformers I and II, we observe that all LTE predicted emission lines of \ce{NH2CH2COOH} are blended and below 2$\sigma$. Based on our LTE modelling, we estimated the 3$\sigma$ upper limit column density of \ce{NH2CH2COOH} conformers I and II were $\leq$3.26$\times$10$^{15}$ cm$^{-2}$ and $\leq$1.20$\times$10$^{13}$ cm$^{-2}$, respectively. The \ce{NH2CH2COOH} conformer I has a dipole moment of $\mu_{a}$ = 0.911 D (a-type) and $\mu_{b}$ = 0.607 D (b-type), whereas conformer II has a dipole moment of $\mu_{a}$ = 5.372 D (a-type) and $\mu_{b}$ = 0.93 D (b-type) \citep{lov95}. The a-type transitions of \ce{NH2CH2COOH} are more evident than the b-type in the ISM because the line intensity of the molecule is proportional to the square of the dipole moment \citep{lov95}. Our successful detection of \ce{CH3NH2} towards the G358.93--0.03 MM1 using ALMA gives more confidence that the G358.93--0.03 MM1 is the ideal candidate for searching for \ce{NH2CH2COOH} and its isomers. We propose a spectral line survey of other possible \ce{NH2CH2COOH} precursor molecules such as \ce{CH2NH}, \ce{NH2CH2CN}, \ce{NH2CH2C(O)NH2}, and also \ce{NH2CH2COOH} conformers I and II with higher angular resolution and higher integration time to solve the puzzle of \ce{NH2CH2COOH} in the ISM.

\section{Conclusion}
\label{conclu}
In this article, we analysed the ALMA band 7 data of the massive star-formation region of G358.93--0.03. The main conclusions of this work are as follows: \\\\
1. We identified a total of sixteen transition lines of \ce{CH3NH2} towards the hot molecular core G358.93--0.03 MM1, which was located in the massive star-formation region G358.93--0.03. Among the sixteen transitions, we observed that only five transitions are non-blended. \\\\
2. The estimated column density of \ce{CH3NH2} towards the G358.93--0.03 MM1 was (1.10$\pm$0.31)$\times$10$^{17}$ cm$^{-2}$ with an excitation temperature of 180.8$\pm$25.5 K. The estimated fractional abundance of \ce{CH3NH2} towards the G358.93--0.03 MM1 with respect to \ce{H2} was (8.80$\pm$2.60)$\times$10$^{-8}$. The column density ratio of \ce{CH3NH2} and \ce{NH2CN} towards G358.93--0.03 MM1 was (1.86$\pm$0.95)$\times$10$^{2}$.\\\\
3. We compared our estimated fractional abundance of \ce{CH3NH2} with the three-phase warm-up model abundance of \ce{CH3NH2}, which was done by \cite{gar13}. After the comparison, we noticed that the fast warm-up model abundance of \ce{CH3NH2} satisfied the observed abundance of \ce{CH3NH2} towards the G358.93--0.03 MM1. That comparison indicates \ce{CH3NH2} may be created on the grain surface via the reaction between radical \ce{CH3} and radical \ce{NH2} towards the G358.93--0.03 MM1. We also compare our estimated abundance of \ce{CH3NH2} towards G358.93--0.03 with other hot molecular core objects G10.47+0.03, Sgr B2, Orion KL, and galactic center cloud G+0.693--0.027. After that comparison, we observed that G358.93--0.03 MM1 is the most highly abundant \ce{CH3NH2} source ever known. We also discuss the possible formation mechanism of \ce{CH3NH2} and the link between the \ce{CH3NH2} and \ce{NH2CH2COOH} in the hot molecular cores. \\\\
4. We also searched the emission lines of the simplest amino acid \ce{NH2CH2COOH} conformers I and II towards the G358.93--0.03 MM1. We did not detect the emission lines of \ce{NH2CH2COOH} conformers I and II within the limits of LTE analysis. The estimated upper limit column densities of glycine conformers I and II were $\leq$3.26$\times$10$^{15}$ cm$^{-2}$ and $\leq$1.20$\times$10$^{13}$ cm$^{-2}$ respectively. \\\\
5. The unsuccessful detection of \ce{NH2CH2COOH} towards G358.93--0.03 MM1 using ALMA indicated that the emission lines of \ce{NH2CH2COOH} may be below the confusion limit in the G358.93--0.03 MM1. Combined spectral line studies of other \ce{NH2CH2COOH} precursor molecules using the LTE model and two or three-phase warm-up chemical modelling are required to understand the prebiotic chemistry towards G358.93--0.03 MM1, which will be carried out in our follow-up study.

\section*{ACKNOWLEDGEMENTS}{
We thank the anonymous rewiewer for her/his constructive comments that helped improve the manuscript. A.M. acknowledges Professor Serena Viti for helping with the discussion part of this manuscript. A.M. acknowledges the Swami Vivekananda Merit-cum-Means Scholarship (SVMCM) for financial support for this research. This paper makes use of the following ALMA data: ADS /JAO.ALMA\#2019.1.00768.S. ALMA is a partnership of ESO (representing its member states), NSF (USA), and NINS (Japan), together with NRC (Canada), MOST and ASIAA (Taiwan), and KASI (Republic of Korea), in co-operation with the Republic of Chile. The Joint ALMA Observatory is operated by ESO, AUI/NRAO, and NAOJ.}

\section*{DATA AVAILABILITY}
The plots within this paper and other findings of this study are available from the corresponding author on reasonable request. The data used in this paper are available in the ALMA Science Archive (\url{https://almascience.nrao.edu/asax/}), under project code 2019.1.00768.S.

\section*{Conflicts of interest}{The authors declare no conflict of interest.}

\section*{CRedit authorship contribution statement}
Arijit Manna: Analysed the ALMA data of G358.93--0.03 and identified the emission lines of \ce{CH3NH2}, Writing original draft. Sabyasachi Pal: Conceptualize the project.

\end{document}